\documentclass[12pt]{article}
\usepackage{amsmath}
\usepackage{amsthm}
\usepackage{amsfonts}
\usepackage{graphicx,psfrag,epsf}
\usepackage{enumitem}
\usepackage{natbib}
\usepackage{url} 
\usepackage{subcaption}
\usepackage{algorithm}
\usepackage{algpseudocode}
\usepackage{float} 
\usepackage{pdflscape} 
\usepackage{booktabs, multirow, multicol, makecell, subcaption}
\usepackage{caption}   
\usepackage{array}
\usepackage{hyperref}
\usepackage{tikz}
\usetikzlibrary{positioning}
\usepackage{makecell}
\usepackage{lipsum, duckuments}
\usepackage{mathtools}

\usepackage{graphicx}

\newcommand{\cut}[1]{{}}

\makeatletter
\newcommand*{\rom}[1]{\expandafter\@slowromancap\romannumeral #1@}
\makeatother

\newcommand{\vX}{{\mathbf{X}}}

\newcommand{\cG}{{\mathcal{G}}}

\newcommand{\cM}{{\mathcal{M}}}

\newcommand{\bbE}{{\mathbb{E}}}

 \usepackage[bb=boondox]{mathalfa}

\usepackage{xparse}
\DeclareFontFamily{U}{ntxmia}{}
\DeclareFontShape{U}{ntxmia}{m}{it}{<-> ntxmia }{}
\DeclareFontShape{U}{ntxmia}{b}{it}{<-> ntxbmia }{}
\DeclareSymbolFont{lettersA}{U}{ntxmia}{m}{it}
\SetSymbolFont{lettersA}{bold}{U}{ntxmia}{b}{it}

\ExplSyntaxOn
\NewDocumentCommand{\varmathbb}{m}
 {
  \tl_map_inline:nn { #1 }
  {
    \use:c { varbb##1 }
  }
 }
\tl_map_inline:nn { ABCDEFGHIJKLMNOPQRSTUVWXYZ }
 {
  \exp_args:Nc \DeclareMathSymbol{varbb#1}{\mathord}{lettersA}{\int_eval:n { `#1+67 }}
 }
\exp_args:Nc \DeclareMathSymbol{varbbk}{\mathord}{lettersA}{169}
\ExplSyntaxOff

\definecolor{lightgrey}{gray}{0.8}
\definecolor{medgrey}{gray}{0.6}
\definecolor{darkgrey}{gray}{0.4}

\newcommand{\blind}{0}

\newtheorem{assumption}{Assumption}

\begin{document}

\def\spacingset#1{\renewcommand{\baselinestretch}%
{#1}\small\normalsize} \spacingset{1}


\if0\blind
{
  \title{\bf A Design-Based Matching Framework for Staggered Adoption with Time-Varying Confounding}
  \author{  Suehyun Kim\thanks{Department of Statistics, Seoul National University} and Kwonsang Lee\footnotemark[1]\\
  \date{}} \maketitle
} \fi

\if1\blind
{
  \bigskip
  \bigskip
  \bigskip
  \begin{center}
    {\LARGE\bf Title}
\end{center}
  \medskip
} \fi

\bigskip
\begin{abstract}
Causal inference in longitudinal datasets has long been challenging due to dynamic treatment adoption and confounding by time-varying covariates. Prior work either fails to account for heterogeneity across treatment adoption cohorts and treatment timings or relies on modeling assumptions. In this paper, we develop a novel design-based framework for inference on group- and time-specific treatment effects in panel data with staggered treatment adoption. We establish identification results for causal effects under this structure and introduce corresponding estimators, together with a block bootstrap procedure for estimating the covariance matrix and testing the homogeneity of group-time treatment effects. To implement the framework in practice, we propose the Reverse-Time Nested Matching algorithm, which constructs matched strata by pairing units from different adoption cohorts in a way that ensures comparability of covariate histories at each treatment time. Applying the algorithm to the Netflix–IPTV dataset, we find that while Netflix subscription does not significantly affect total IPTV viewing time, it does negatively affect VoD usage. We also provide statistical evidence that the causal effects of Netflix subscription may vary even within the same treatment cohort or across the same outcome and event times.

\end{abstract}

\noindent%
{\it Keywords:} Design-based inference; Nested structure; Matching algorithm; Panel data

\vfill

\newpage
\spacingset{1.45} 

\section{Introduction}
\label{sec1}

\subsection{Motivation}
\label{sec1.1}
Over the past decade, the rapid expansion of over-the-top (OTT) media services has significantly transformed audiovisual consumption patterns. Global streaming platforms, including Netflix, Disney+, and YouTube, have experienced substantial growth driven by extensive on-demand content libraries, cross-device accessibility, and personalized recommendation algorithms that leverage user-level behavioral data. This growth accelerated during the COVID-19 pandemic and was accompanied by a marked decline in IPTV usage, indicating that subscription-based streaming platforms are increasingly serving as substitutes for real-time television and IPTV-based video-on-demand (VoD) services \citep{jung2024iptv}.

Nevertheless, OTT adoption exhibits considerable heterogeneity. The timing and duration of subscriptions vary widely among individuals, and distinct subscriber cohorts may interact with platforms in diverse ways. For example, the release of \textit{Squid Game} on Netflix in September 2021 resulted in an unprecedented surge in new subscribers, driven by a single, highly prominent program that remains the platform’s most-watched series as of 2025. These shock-induced entrants, who subscribed in response to a blockbuster content release, may systematically differ from earlier adopters in their viewing behaviors, responsiveness to personalized recommendations, and opportunity cost of time.

Motivated by these observations, we address two empirical questions: (1) whether OTT adoption causally affects IPTV viewing behavior, and (2) whether the magnitude and dynamics of this causal effect differ according to subscription timing and duration. Using longitudinal user-level data from a Korean broadcasting company, we examine the causal impact of Netflix subscription on total viewing time and VoD viewing time. The panel dataset is recorded at the monthly level and consists of a randomly selected sample of the company’s IPTV subscribers. For each household, we observe repeated measures of Netflix usage status, real-time IPTV viewing hours, and IPTV-based VoD viewing hours. A key methodological challenge for causal inference in this setting is the staggered timing of treatment initiation, as each unit subscribes to Netflix at different times, which may reflect time-varying confounders, especially prior viewing patterns.

In this work, we present a design-based framework to address time-varying confounding in longitudinal datasets and enable simultaneous inference on treatment effects by timing and duration. We introduce a novel matching procedure, \textit{Reverse-Time Nested Matching}, which constructs comparable matched strata across multiple adoption cohorts under the staggered treatment adoption regime. This approach emulates a randomized experiment at each adoption time, in line with the broader design-based tradition of randomization inference \citep{rosenbaum2002observational}, and is straightforward to implement using optimal full matching. Applying this framework to our Netflix-IPTV dataset, we find that Netflix subscription causally increases VoD viewing time but has no significant effect on total viewing time. Moreover, the magnitude of the effect varies across adoption cohorts even when treatment duration is constant.

\subsection{Related literature and contributions}
\label{sec1.2}

Longitudinal observational studies frequently involve staggered treatment initiation and their confounding due to time-varying covariates, a scenario particularly prevalent in clinical settings where patients self-select into treatment \citep{hade2020psm}. A prominent design-based approach in this context is \emph{risk set matching}, first introduced by \citet{li2001riskset} and subsequently extended through propensity-score formulations \citep{lu2005psm, hade2020psm}. Risk set matching pairs units with similar covariate trajectories at the time of treatment initiation, without conditioning on post-treatment events \citep{lu2023handbook}, thereby approximating an ideal randomized experiment with respect to the risk sets.

More recently in the matching literature, \citet{imai2023panelmatch} proposed \emph{PanelMatch}, a matching method for time-series cross-sectional data that constructs matched sets based on treatment histories and subsequently performs matching with replacement on covariates. While PanelMatch is not intended to emulate a blocked randomized experiment, it employs matching as a nonparametric adjustment within a difference-in-differences framework to estimate average treatment effects.

Both approaches, however, face limitations when treatment timing and duration generate heterogeneity in treatment effects. Risk set matching aggregates matches across varying adoption times, implicitly averaging over heterogeneous treatment timings and future outcome trajectories. In contrast, PanelMatch conditions on treatment histories but does not isolate variation attributable to the timing of initial treatment.

On the other hand, \citet{callaway2021did} developed a difference-in-differences framework for group-time average treatment effects that incorporates both treatment timing and cohort structure in staggered adoption regimes, along with aggregated causal parameters obtained by averaging the group-time effects across cohorts and time periods. However, the associated regression- and weighting-based estimators remain model dependent. Furthermore, while group-time treatment parameters can illustrate heterogeneity across treatment timing and cohorts, the estimation and inference procedure does not formally assess such heterogeneity.

In light of these limitations, our work makes the following contributions:

\begin{enumerate}[label = (\roman*)]
    \item We propose a new matching procedure for longitudinal data with staggered treatment adoption and time-varying covariates. The method constructs matched strata that include units from all adoption cohorts, while ensuring comparability in covariate histories at any treatment time of interest. The resulting design emulates a localized randomized experiment at each adoption time, where accumulating time-varying covariates naturally produce progressively finer strata. Such a structure yields an alternative estimator for the group-time average treatment effects of \citet{callaway2021did}.
    \item We develop an inference procedure based on a block bootstrap that estimates the covariance matrix of multiple group time-treatment effects, accounting for correlations across groups and time periods. This enables simultaneous hypothesis testing for whether treatment effects are equal when (i) the adoption cohort is fixed, (ii) the outcome is measured at the same time, or (iii) treatment duration is held constant. These tests allow a formal assessment of treatment effect heterogeneity across adoption cohorts and treatment timing.
    \item The matching method is straightforward to implement using optimal full matching \citep{rosenbaum1991optimal, hansen2004ofm}, and retains the main advantages of matching methods in observational studies, including reduced model dependence, simple interpretation, and absence of outcome information in the design  \citep{ho2007matching}. Moreover, our estimators rely on a cross-sectional ignorability assumption for time-varying covariates, relaxing the stronger parallel trends assumptions required in DiD-based frameworks \citep{callaway2021did, caetano2024did1, caetano2024did2}.
\end{enumerate}

The remainder of the paper is organized as follows. Section \ref{sec2} introduces the causal framework and notation for longitudinal settings and presents identification results for group-time average treatment effects. Section \ref{sec3} details the proposed matching algorithm and the corresponding estimators, and develops a bootstrap-based inference procedure along with a hypothesis test for treatment effect homogeneity. In Section \ref{sec4}, we apply the method to the Netflix–IPTV dataset and discuss the empirical findings. Section \ref{sec5} concludes with key takeaways and directions for future research. Proofs, additional analyses, and supplementary figures are provided in the Supplementary Materials.

\section{Causal framework in longitudinal observational studies}
\label{sec2}

\subsection{Notation and setup}
\label{sec2.1}

To formalize the problem, we first introduce the notation used throughout the paper. Consider a balanced panel of $N$ units observed over discrete time periods $t = T_0, \dots, 0, 1, \dots, T$. The period $t = 1$ marks the earliest possible treatment, while $t = T_0, \dots, 0$ correspond to pre-treatment periods in which time-varying confounders may be measured. For each unit $i = 1, \dots, N$, we observe the longitudinal data $(Y_{i1}, \dots, Y_{iT}, Z_{i1}, \dots, Z_{iT}, \vX_{iT_0}, \dots, \vX_{iT})$, where $Y_{it}$ and $\vX_{it}$ denote the observed outcome and covariates of unit $i$ at time $t$, and $Z_{it}$ is a binary indicator of treatment status. Although treatment timing may differ across units, we impose the following \textit{staggered adoption} structure:

\begin{assumption}[Staggered treatment adoption]\label{assm:staggered}
    For $t = T_0, \dots, 0$, $Z_t = 0$. For $t = 2, \dots, T$, $Z_{t-1} = 1$ implies $Z_t = 1$.
\end{assumption}

In words, once a unit receives treatment, it remains treated in subsequent periods. This can be viewed as an irreversibility condition, reflecting that units do not revert to the untreated state or forget prior treatment exposure \citep{callaway2021did}.

Let $G_i \in \{1, 2, \dots, T, \infty\}$ denote the period in which unit $i$ first receives treatment, with $G_i = \infty$ indicating that unit $i$ is never treated during $t = 1, \dots, T$. Under Assumption \ref{assm:staggered}, the treatment path $(Z_{i1}, \dots, Z_{iT})$ is fully determined by $G_i$. We refer to $\{i : G_i = g\}$ as the \textit{cohort} (or \textit{group}) initiating treatment in period $g$. To define causal parameters, we adopt the potential outcomes framework \citep{rubin1974estimating}. Following \cite{callaway2021did}, potential outcomes are indexed by calendar time $t$ and hypothetical adoption time $g$: $Y_{it}(g)$ represents the outcome unit $i$ would realize at time $t$ if it began treatment in period $g$ and remained treated thereafter, while $Y_{it}(\infty)$ denotes the outcome that would be observed if the unit never received treatment during the study period.

A fundamental causal quantity in staggered treatment adoption settings is the \textit{group-time average treatment effect}, denoted by $ATT(g,t)$. For $g = 1, \dots, T$ and $t = 1, \dots, T$, we define
\begin{equation*}
    ATT(g, t) \coloneq \bbE [Y_{it}(g) - Y_{it}(\infty) \mid G_i = g].
\end{equation*}

Here, $ATT(g, t)$ represents the treatment effect at time $t$ for the cohort $\{i : G_i = g\}$ that first received treatment in period $g$. Such a formulation is convenient as it explicitly distinguishes between calendar time $t$ and cohort $g$, allowing for a clear characterization of heterogeneity across both dimensions.

\subsection{Identification of the group-time average treatment effect}
\label{sec2.2}

To identify the group-time average treatment effects $ATT(g, t)$, we impose the following set of assumptions.

\begin{assumption}[Stable unit treatment value assumption (SUTVA)] \label{assm:sutva}
    For all $i = 1, \dots, N$ and $t \in \{1, \dots, T, \infty\}$, we have $Y_{it} = Y_{it}(G_i)$, and $Y_{it}(G)$ does not depend on any other units $j$ such that $i \neq j$.
\end{assumption}

\begin{assumption}[Stable pre-treatment outcomes] \label{assm:stable}
    For all $g \in \{1, \dots, T, \infty\}$ and $t < g$, $Y_{it}(g) = Y_{it}(\infty)$.
\end{assumption}

Assumption \ref{assm:sutva} represents the standard consistency and no interference condition, stating that the observed outcome equals the potential outcome corresponding to the unit’s realized treatment history. Assumption \ref{assm:stable} ensures that the treatment has no impact on outcomes prior to exposure. We also impose the following assumptions regarding potential confounders.

\begin{assumption}[Time-specific unconfoundedness]
\label{assm:tsu}
For each cohort $g = 1, \dots, T$ and potential treatment adoption time $t \geq g$,
    \begin{equation*}
        (Y_{it}(g), Y_{it}(\infty)) \; \perp \; G_i \mid (\vX_{iT_0}, \dots \vX_{i, t-1}).
    \end{equation*}
\end{assumption}

\begin{assumption}[Overlap]
\label{assm:overlap}
    For each $g \in \{1, \dots, T, \infty \}$,
    \begin{equation*}
        0 < \Pr(G_i = g \mid \vX_{iT_0}, \dots, \vX_{iT} ) < 1 \text{ and } 0 < \Pr(G_i > t \mid \vX_{iT_0}, \dots, \vX_{iT} ) < 1.
    \end{equation*}
\end{assumption}

Assumption \ref{assm:tsu} states that for any given time period, the timing of treatment adoption is as good as random conditional on the observed covariates up to period $t-1$. Assumption \ref{assm:overlap} ensures that, no cohort is deterministically ruled out for any possible value of the observed covariates. Under Assumptions \ref{assm:staggered}-\ref{assm:overlap}, the group-time average treatment effect $ATT(g, t)$ is identified as
\begin{equation*}
ATT(g,t)
=
\mathbb{E}\!\left[
  \mathbb{E}[Y_{it}\mid G_i=g,\vX_{i,t-1}]
 -
  \mathbb{E}[Y_{it}\mid G_i>t,\vX_{i,t-1}]
\right].
\end{equation*}

Note that this set of assumptions provides an alternative identification strategy for the $ATT(g,t)$s, primarily through Assumption~\ref{assm:tsu}. This assumption imposes cross-sectional exchangeability at each time point $t$, rather than the conditional parallel trends assumption used in DiD-based frameworks \citep{callaway2021did, caetano2024did1, caetano2024did2}. Such an assumption is also more compatible with the design-based approach introduced in the next section.

\subsection{Interpretation of the group-time average treatment effects}
\label{sec2.3}

Because staggered treatment adoption generates heterogeneity along both calendar time and exposure duration, it is useful to organize the $ATT(g,t)$ estimands along specific dimensions that facilitates interpretation and empirical analysis. In the following, we outline three complementary perspectives that summarize how treatment effects vary across cohorts, over time, and by exposure length.

First, fixing the adoption cohort $g$, the sequence
\begin{equation*}
     \{ ATT(g,t) : t = g, g{+}1, \dots, T \}
\end{equation*}
captures the \emph{cohort-wise dynamic effect profile}. This trajectory describes how the causal effects evolves as time elapses after treatment adoption in period $g$. Such profiles may contain information on persistence, decay, or delayed onset of treatment effects, and are often the primary object of interest in longitudinal evaluations.

Second, fixing a calendar time $t$, the set
\begin{equation*}
    \{ ATT(g,t) : g \leq t \}
\end{equation*}
summarizes \emph{cross-cohort heterogeneity} in the effect at time $t$. Such a perspective highlights how the magnitude of the effect at a given point in time depends on treatment initiation, enabling comparisons between early and late adopters. 

Finally, treatment effects can be examined in terms of exposure duration. Let $e = t - g \ge 0$ denote the \emph{event time} or length of exposure. Then the set
\begin{equation*}
    \{ ATT(g, g + e) : g = 1, 2, \dots, T - e \}
\end{equation*}
represents the \emph{fixed lag effects}, that is, the causal effect $e$ periods after treatment initiation in all eligible cohorts. These effects isolate the relationship between exposure length and treatment response from a specific calendar time. Examining whether $ATT(g,g{+}e)$ varies with $g$ provides a direct assessment of whether the $e$-lag effect depends on the timing of treatment adoption.

\section{Matching algorithm}
\label{sec3}

\subsection{Reverse-time nested matching (RTNM)}
\label{sec3.1}

In this section, we develop a novel matching algorithm for the estimation and inference of $ATT(g,t)$. Due to the accumulation of time-varying confounders over time, the estimation of $ATT(g,t)$ for different cohorts $g$ requires adjustment for different subsets of covariates. For example, estimating $\widehat{ATT}(3,t)$ requires adjustment for $(\vX_{T_0}, \dots, \vX_{2})$, whereas estimating $\widehat{ATT}(4,t)$ requires adjustment for $(\vX_{T_0}, \dots, \vX_{3})$. More generally, the identification of $ATT(g,t)$ requires conditioning on
\begin{equation*}
\vX_{T_0:(g-1)} \coloneq (\vX_{T_0}, \dots, \vX_{g-1}),
\end{equation*}
so that if $g < g'$, then $\vX_{T_0:(g-1)} \subset \vX_{T_0:(g'-1)}$. This induces a natural nested structure in the required covariate adjustment sets across cohorts.

\begin{figure}[h!]
    \centering
    \includegraphics[width=0.85\textwidth]{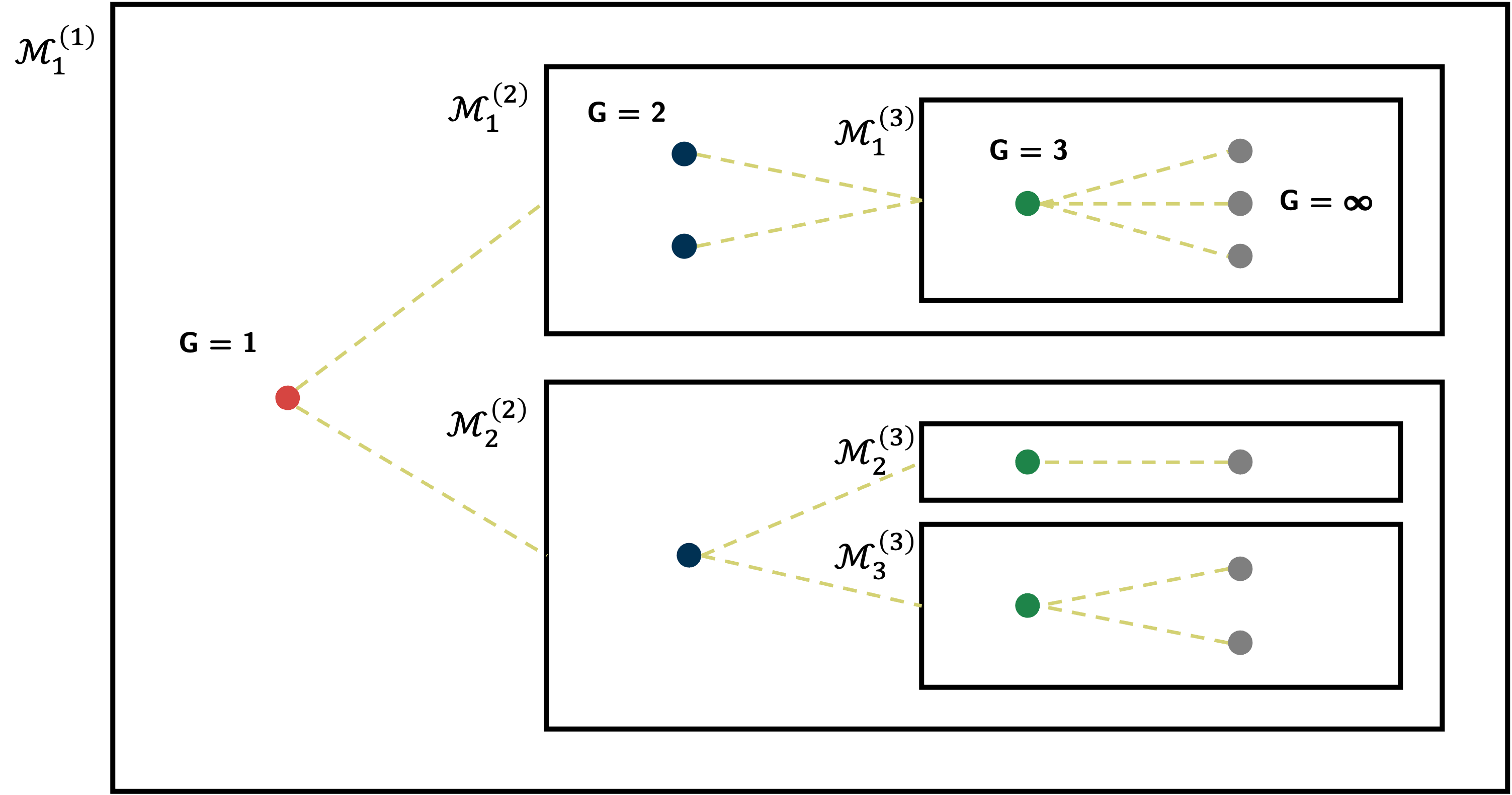}
    \caption{Example of a nested design with time-varying covariates with 4 treatment adoption cohorts. The red, blue, green, and gray points correspond to units from cohorts $G = 1, 2, 3, \infty$, respectively. Units within the same boxes exhibit comparable covariate values in the periods prior to treatment adoption. The dashed lines indicate full matches in the RTNM algorithm, where strata from the later treated period are treated as a single pseudo-control unit.}
    \label{fig:nested}
\end{figure}

We propose a novel matching procedure, the \textit{Reverse-Time Matching Algorithm (RTNM)}, that leverages this structure to enable simultaneous estimation of $ATT(g,t)$ across all cohorts and post-treatment periods. First, we present relevant notations. Let $\cM^{(g)}_j$, $j = 1, \dots, n_g$ denote the $j$-th matched set that contains treated and control units with respect to the cohort $g$, i.e. units first treated at time period $g$ and not-yet-treated units up to $g$. Here, we assume that we have a fixed distance metric $d$ that measures the discrepancy between the covariates to form the matched sets, such as the propensity score distance or the Mahalanobis distance. Let $d_g(i, j) \coloneq d(X_{i, T_0:(g-1)}, X_{j, T_0:(g-1)})$ denote the distance between units $i$ and $j$ with respect to the time period $g$, so that it measures the distance between pre-$g$ treatment covariates. We generalize the distance metric so that it can measure distances between a single unit and a matched set. Let $\cM^{(g^*)}_j$ be a matched set. For $g < g^*$ and units $i$ in cohort $g$, define
\begin{equation}\label{distance}
    d_g(i, \cM^{(g^*)}_j) \coloneq \frac{1}{|\cM^{(g^*)}_j|} \sum_{k \in \cM^{(g^*)}_j} d_g(i, k)
\end{equation}
so that $d(i, \cM^{(g^*)}_j)$ is the average distance from unit $i$ to units in $\cM^{(g^*)}_j$ based on covariates $\vX_{T_0:(g-1)}$.

The goal of the algorithm is to create matched sets of units from each cohort $g = 1, \dots, G$, which exhibits a nested structure that ensures covariate adjustments for any time point, as shown in Figure \ref{fig:nested}. In order to do so, the RTNM algorithm constructs matched sets beginning from the latest time period of interest and proceeds backward. Because the covariate adjustment set gets finer as time proceeds, matching in reversed-order ensures that each subsequent iteration refines the adjustment set in a way that respects the nested structure. At each step, the matched set from the previous iteration is regarded as a \textit{pseudo-control} unit and is matched to a treated unit using the covariates appropriate for that time point. This procedure guarantees the required nested structure of covariate adjustments while producing valid matched sets for the estimation of all $ATT(g,t)$. The steps of the matching procedure are formally outlined below. We assume that full matching is implemented at each step, so that no units are discarded during the matching process.

\paragraph{Step 1 (Initial step).} Optimally match units from cohort $\{i: G_i = G\}$ to $\{i: G_i > G\}$ to form matched sets $\cM^{(G)}_1, \dots, \cM^{(G)}_{n_G}$. 

\paragraph{Step 2 (Matching with pseudo-controls).} Move to the previous cohort $G-1$, and match units from cohort $\{i: G_i = G - 1\}$ to matched sets $\cM^{(G)}_1, \dots, \cM^{(G)}_{n_G}$ as follows:
\begin{enumerate}[label = (\roman*)]
    \item Compute the distance matrix between the treated cohort $\{i: G_i = G -1\}$ and the not-yet-treated cohort $\{i: G_i > G -1 \}$, based on the prespecified metric $d$.
    \item Based on the distance matrix from (i), compute the distance from each unit in the treated cohort $\{i: G_i = G -1\}$ to the matched sets $\cM^{(G)}_1, \dots, \cM^{(G)}_{n_G}$, using the metric given by Equation \ref{distance}.
    \item Using this distance, optimally match units in $\{i: G_i = G -1\}$ to $\cM^{(G)}_1, \dots, \cM^{(G)}_{n_G}$ to obtain matched sets $\cM^{(G-1)}_1, \dots, \cM^{(G-1)}_{n_{G-1}}$.
\end{enumerate}

\paragraph{Step 3 (Iteration).} Repeat Steps 1-2 to match $\{i: G_i = g\}$ with the previously-matched $\cM^{(g+1)}_1, \dots, \cM^{(g+1)}_{n_{g+1}}$ until the first cohort $
\{i: G_i = 1\}$ is reached.

\bigskip

Consequently, we obtain matched sets $\cM^{(g)}_j$, $j = 1, \dots, n_g$ for each $g = 1, \dots, G$ with the following properties. First, each $\cM^{(g)}_j$ contains at least one unit from cohort $g^*$ such that $g^* \geq g$. In particular, $\cM^{(1)}_j$ contains at least one unit from all cohorts of interest. Second, the matched sets form a nested structure, in the sense that for each $g = 2, \dots, G$ and any matched set $\cM^{(g)}_j$, there uniquely exists a matched set $\cM^{(g-1)}_k$ such that $\cM^{(g)}_j \subset \cM^{(g-1)}_k$. Moreover, for any given time point $t$, the cohorts $\{i: G_i = t \}$ and $\{i: G_i > t\}$ are comparable with respect to the covariates up to past covariate histories $\vX_{T_0 : (t-1)}$. This is due to the optimal matching based on the distance metric $d_g$, which is updated at each iteration so that it reflects the closeness of the strata up to the relevant time point. The dashed lines in Figure \ref{fig:nested} illustrate how the nested design can be constructed using the RTNM algorithm.


\subsection{Inference and simultaneous hypothesis testing under RTNM}
\label{sec3.2}

The RTNM algorithm partitions the sample into disjoint, nested matched blocks, which allows a simultaneous estimation of multiple $ATT(g,t)$ values from a single matched design. Building on this attractive property, we propose a bootstrap procedure to estimate the covariance matrix of the $ATT(g,t)$ estimands of interest. The procedure follows is based on a block-bootstrap logic \citep{lahiri2003resampling}, where we resample the outermost nested blocks that preserve the full nested structure across periods. Each nested block is treated as an independent sampling unit, reflecting the without-replacement design over time.

The block bootstrap for RTNM is implemented as follows. Suppose that we run $B$ bootstrap iterations, and let $\cG = \{(g_1,t_1),\dots,(g_K,t_K)\}$ denote the collection of group--time pairs for which we seek to estimate $ATT(g,t)$. Suppose that the RTNM algorithm has been applied to the sample, and let $n_1$ be the number of matched sets associated with the first adoption cohort $G = 1$, which forms the outermost set of nested blocks. Let $\tau = (ATT(g_1,t_1), \dots, ATT(g_K,t_K))$ be the vector of group--time average treatment effects of interest, and let $\hat{\tau} = (\widehat{ATT}(g_1,t_1), \dots, \widehat{ATT}(g_K,t_K))$ denote the corresponding RTNM estimator. For each outermost matched block $\cM^{(1)}_m$, $m = 1,\dots,n_1$, let $\hat{\tau}_m$ denote its contribution, together with all nested blocks contained in $\cM^{(1)}_m$, to the estimator $\hat{\tau}$.

\paragraph{Step 1.}
Let $\{\widehat{\tau}_m\}_{m=1}^{n_G}$ be the block-level effect estimates. For each bootstrap iteration $b=1,\ldots,B$, draw $n_1$ indices with replacement from $\{1,\dots,n_1\}$ to form the index set $\mathcal{I}^{*(b)}$, and compute
\begin{equation*}
\widehat{\tau}^{*(b)}
= \frac{1}{n_1} \sum_{m\in \mathcal{I}^{*(b)}} \widehat{\tau}_m.
\end{equation*}

\paragraph{Step 2.}
Let
\begin{equation*}
\widehat{  \tau}^{*(b)}
=
\bigl(
\widehat{ATT}^{*(b)}(g_1, t_1),
\dots,
\widehat{ATT}^{*(b)}(g_K,t_K)
\bigr)
\end{equation*}
be the vector of bootstrap estimates of $\tau$ from the $b$-th bootstrap iteration. The bootstrap covariance matrix $\hat \Sigma$ is obtained by computing the sample covariance matrix of $\widehat{  \tau}^{*(b)}$, $b = 1, \dots, B$.

The resulting covariance matrix can be used to construct marginal confidence intervals for individual $ATT(g,t)$ values, as well as to conduct hypothesis tests involving the full causal estimand $\tau$. Of particular interest are tests of homogeneity for a selected subset of group-time effects, formulated under the null hypothesis
\begin{equation*}
    H_0: ATT(g_{i_1},t_{i_1}) = \dots = ATT(g_{i_M}, t_{i_M}),
\end{equation*}
which assesses whether a subvector of $\tau$ is constant across the specified indices.

More generally, such hypotheses can be written in the form $H_0: R\tau = 0$, where $R$ is a projection matrix that projects the estimand of interest onto the null space. Under this formulation, we can construct a null-restricted bootstrap test that approximates the sampling distribution of the estimator under the imposed constraints \citep{mackinnon2009bootstrap, cavaliere2017bootstrap}. More specifically, we can compute the null-restricted estimator
\[
\widehat{  \tau}_0
=
\widehat{  \tau}
- R^{\!\top}\!\bigl(R R^{\!\top}\bigr)^{-1} R\widehat{  \tau}
\]
is obtained by projecting $\widehat{  \tau}$ onto the null space $\{\tau: R  \tau = 0\}$. For each bootstrap draw $\widehat{  \tau}^{*(b)}$, we draw the null-centered version
\[
\widehat{  \tau}^{*(b)}_{\text{null}}
=
\widehat{  \tau}_0
+ \bigl(\widehat{  \tau}^{*(b)} - \widehat{  \tau}\bigr).
\]

Inference is based on the comparison between the observed Wald statistic
\begin{equation*}
W_{\mathrm{obs}}
=
(R\widehat{  \tau})^{\!\top}
  (R\widehat{\Sigma}R^{\!\top})^{-1}
(R\widehat{  \tau}),
\end{equation*}
and the corresponding bootstrap statistics
\begin{equation*}
W^{*(b)}
=
\bigl(R\widehat{  \tau}^{*(b)}_{\text{null}}\bigr)^{\!\top}
  (R\widehat{\Sigma}R^{\!\top})^{-1}
\bigl(R\widehat{  \tau}^{*(b)}_{\text{null}}\bigr).
\end{equation*}
Consequently, the bootstrap $p$-value is computed as $\widehat{p}_B
=
\frac{1}{B}\sum_{b=1}^B 
\mathbf{1}\!\left\{ W^{*(b)} \ge W_{\mathrm{obs}} \right\}$, where $B$ is the number of bootstrap repetitions. Since $W_{\mathrm{obs}}/\operatorname{rank}(R)$ is asymptotically equivalent to an $F$ statistic, this yields an $F$-type test for homogeneity of the selected group–time effects.

In practice, the analyst would conduct two separate bootstrap procedures: one to estimate the covariance matrix $\widehat{\Sigma}$ and another to perform hypothesis testing using this previously obtained estimate of $\Sigma$. These two resampling steps are independent, and the number of bootstrap replications $B$ used for estimating $\widehat{\Sigma}$ need not coincide with the number used for the hypothesis test.

\section{Data application}
\label{sec4}

\subsection{Group-time treatment effects in the Netflix-IPTV dataset}
\label{sec4.1}

We now apply our framework and methodology to the Netflix-IPTV dataset, provided by a Korean IPTV operator firm, in order to address the motivating questions. Because IPTV services are bundled with broadband internet, the provider observes household-level Netflix access through internet usage logs. The sample consists of IPTV subscribers randomly selected at the household level and observed monthly from March 2021 to November 2021, forming a balanced household-level panel. In particular, this period covers the release of the \emph{Squid Game}, which was in September 2021. 

For each household and month, the dataset records Netflix subscription status; total and genre-specific real-time TV viewing hours; total and genre-specific VoD viewing hours; and total VoD purchase expenditures. The treatment variable is an indicator for a paid Netflix subscription with observable activity during the month. Our analyses focus on three outcomes: total real-time viewing hours, total VoD viewing hours, and a binary indicator for any VoD viewing in that month. This binary indicator for VoD viewing addresses the zero-inflated and skewed distribution of monthly VoD viewing hours.

To adjust for time-varying confounding due to self-selection into Netflix adoption, we consider covariates regarding viewing histories. Following \citet{jung2024iptv}, we use nine time-varying covariates: total real-time viewing hours; genre-specific real-time viewing hours for movies and entertainment; total VoD viewing hours; genre-specific VoD viewing hours for movies, dramas, and entertainment; an indicator for any VoD viewing in the past three months; and an indicator for any VoD purchases in the past three months. Notably, total real-time and VoD viewing hours serve both as outcomes of interest and as predictors of future Netflix adoption, creating a setting in which lagged outcomes act as confounders.

Of the nine months of the observation period, we use the first three months (March to May) exclusively for covariate adjustment and restrict the analysis to units whose first Netflix adoption occurs in June or later. For notational convenience, we index the first three months as pre-treatment periods and set $T_0 = -2$, so that these months contribute only covariate information. We take June as the first period in which treatment effects are evaluated and index it by $t = 1$, and the corresponding adoption cohort is denoted $G = 1$. Our analysis considers adoption cohorts $G = 1, 2, 3, 4$, and uses the final two months of data to compute group-time average treatment effects with $t \ge g$.

\begin{figure}[!ht]
    \centering

    \begin{subfigure}{0.9\textwidth}
        \centering
        \includegraphics[width=\textwidth]{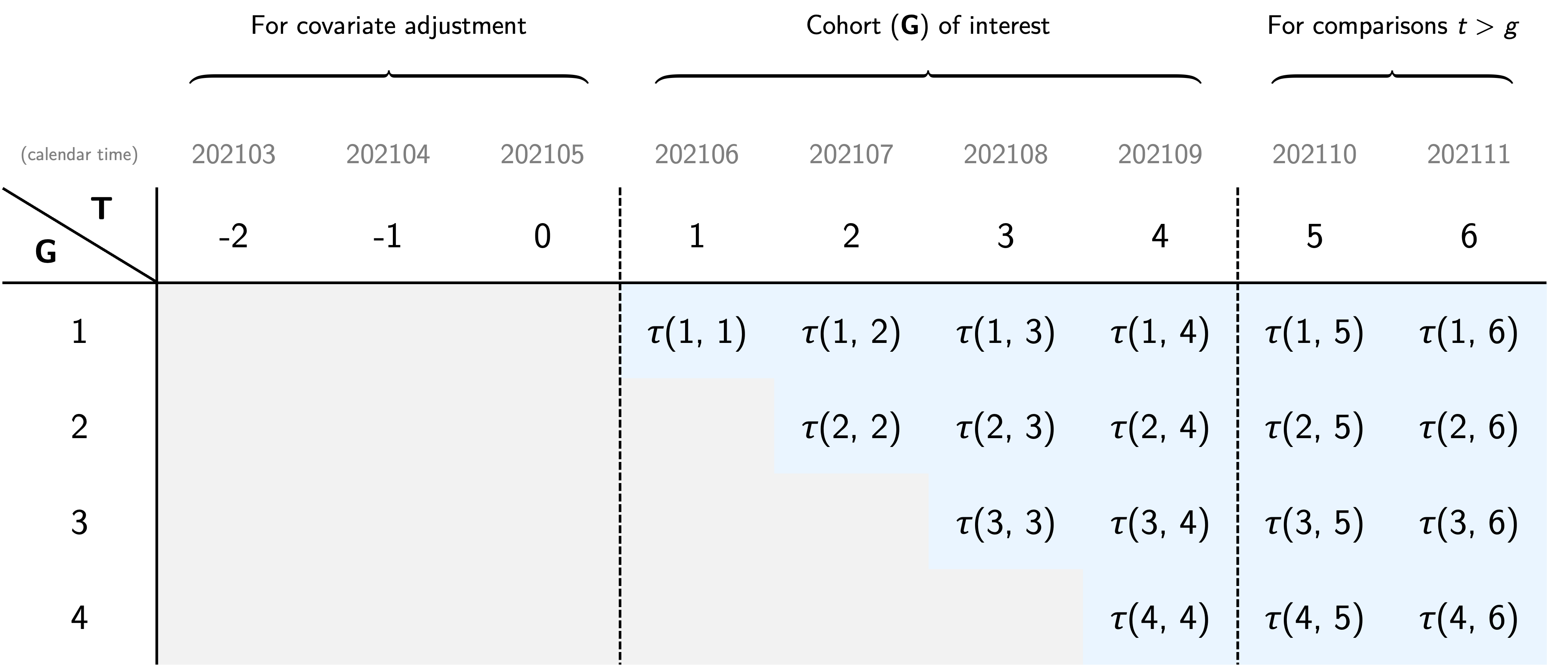} 
        \caption{Target causal estimands $ATT(g,t)$ in the Netflix-IPTV dataset. The rows and columns, denoted by $G$ and $T$, indicate the calendar time and the treatment adoption cohorts, respectively.}
        \label{fig:structure_all}
    \end{subfigure}
    \hfill

    \begin{subfigure}{0.9\textwidth}
        \centering
        \includegraphics[width=\textwidth]{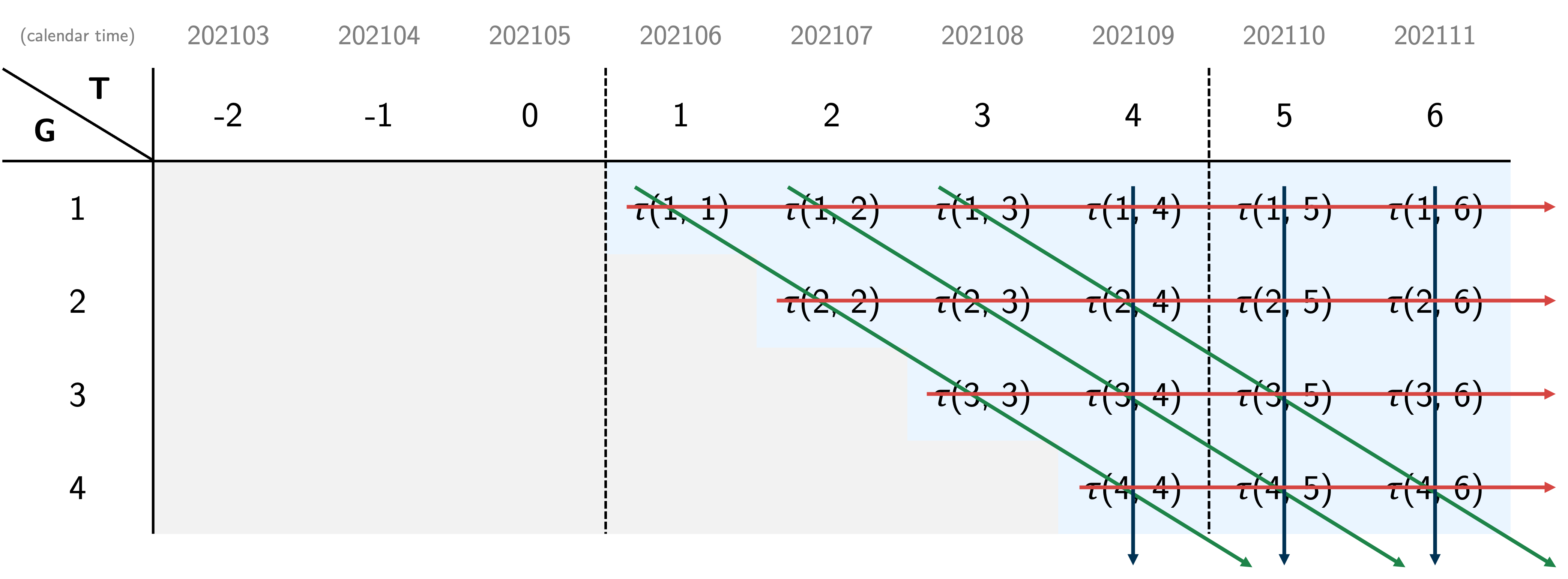} 
        \caption{Hypothesis test for the homogeneity of $ATT(g,t)$ across different dimensions. Red, blue, and green arrows indicate tests for fixed cohort, fixed calendar time, and fixed lag, respectively.}
        \label{fig:structure_hypo}
    \end{subfigure}

    \caption{Causal estimands and hypothesis tests of interest.}
    \label{fig:structure}
\end{figure}

Under this setup, group-time average treatment effects are defined for all $(g,t)$ pairs with $g = 1, 2, 3, 4$ and $g \le t \le 6$, yielding a total of 18 group-time effects $ATT(g,t)$. The structure is visualized in Figure \ref{fig:structure_all}. After restricting the sample to households that adopt treatment from June onward, the analysis includes 9{,}627 units in total. Among the sample, the numbers of units in cohorts $G = 1, 2, 3, 4$ are 237, 360, 302, and 838, respectively, and 7{,}890 units are never treated during the observation window.

Within this structure, we implement Reverse-Time Nested Matching to construct matched strata consisting of comparable units from each cohort $G = 1, 2, 3, 4, \infty$. Using the \texttt{R} package \texttt{MatchIt}, which implements optimal full matching via the \texttt{optmatch} package, we employ the rank-based Mahalanobis distance as the matching metric to reduce sensitivity to outliers and binary covariates, following the recommendations of \citet{rosenbaum2020dos}. Across treated and control groups for each cohort $G = 1,2,3,4$, most covariates achieve an absolute standardized mean difference below 0.1, indicating satisfactory covariate balance and comparable to the level of finite-sample balance obtained from marginal optimal full matching at each time point. Additional matching diagnostics and comparisons with marginal optimal full matching are presented in the Supplementary Materials. Also, note that the choice of the maximum stratum size may affect the resulting matches \citep{stuart2010matching}; refer to the Supplementary Materials for a detailed discussion of this hyperparameter selection.

After constructing the matched stratified design using RTNM, we estimate the treatment effects using the matched structure with regression adjustment \citep{ho2007matching, stuart2010matching}. Standard errors for the point estimates are computed using the bootstrap procedure described in Section~\ref{sec3.2}, with $5{,}000$ bootstrap repetitions to estimate the covariance matrix of all $ATT(g,t)$ values of interest.

The results for the three outcome variables are presented in Table \ref{tab:outcomes}. Comparing Table \ref{tab:outcome1} with Tables \ref{tab:outcome2} and \ref{tab:outcome3}, we observe that none of the $ATT(g,t)$ estimates for total real-time viewing hours are statistically significant at the $\alpha = 0.05$ level, whereas numerous significant effects appear for VoD viewing behavior. Moreover, all point estimates for total VoD viewing hours and VoD viewing status are negative. These patterns suggest that while Netflix subscription may not influence overall media consumption, it may reduce VoD usage, indicating that Netflix likely serves as a substitute for IPTV-based VoD services.

On the other hand, VoD viewing behavior exhibits substantial heterogeneity between cohorts. For instance, the cohort $G = 2$, which initiated treatment in July, shows a strong negative group-time effect for both total VoD viewing hours and VoD viewing status, whereas for the cohort $G = 3$, which adopted treatment in August, the negative causal effects are noticeably weaker. We formally assess such heterogeneity through hypothesis tests in the next subsection.

\begin{table}[!ht]
    \centering
    \begin{subtable}{\textwidth}
        \centering
        \footnotesize
        
        \begin{tabular}{c cccccc}
            \hline
            $(g,t)$ & 1 & 2 & 3 & 4 & 5 & 6 \\ 
            \hline
            1 & -1.635 (5.881) & -0.145 (6.514) & -1.603 (5.971) & 2.026 (5.927) & -9.733 (5.786) & -7.500 (5.759) \\ 
            2 & - & -3.561 (4.276) & -5.389 (5.086) & 0.819 (5.161) & -1.149 (5.530) & -9.254 (4.973) \\ 
            3 & - & - & 5.232 (5.087) & -0.851 (5.166) & -1.691 (5.083) & -0.590 (6.154) \\ 
            4 & - & - & - & 1.613 (2.600) & -3.428 (3.334) & 2.388 (3.510) \\ 
            \hline
        \end{tabular}
        \caption{Total real-time viewing hours}
        \label{tab:outcome1}
    \end{subtable}

    \vspace{1em}

    \begin{subtable}{\textwidth}
        \centering
        \footnotesize
        
        \begin{tabular}{c cccccc}
            \hline
            $(g,t)$ & 1 & 2 & 3 & 4 & 5 & 6 \\ 
            \hline
            1 & -1.281 (1.189) & \textbf{-4.453 (1.492)} & \textbf{-4.236 (1.456)} & \textbf{-3.443 (1.498)} & \textbf{-3.286 (1.492)} & \textbf{-3.207 (1.376)} \\ 
            2 & - & \textbf{-4.346 (0.985)} & \textbf{-4.000 (1.111)} & \textbf{-3.729 (1.049)} & \textbf{-4.823 (0.844)} & \textbf{-2.371 (0.948)} \\ 
            3 & - & - & \textbf{-2.900 (0.905)} & -1.782 (0.966) & -0.716 (1.239) & -1.183 (1.189) \\ 
            4 & - & - & - & \textbf{-1.641 (0.587)} & \textbf{-1.647 (0.625)} & \textbf{-2.259 (0.685)} \\ 
            \hline
        \end{tabular}
        \caption{Total VoD viewing hours}
        \label{tab:outcome2}
    \end{subtable}

    \vspace{1em}

    \begin{subtable}{0.8\paperwidth}
    \centering
    \footnotesize
    \begin{tabular}{c ccccccc}
        \hline
        $(g,t)$ & 1 & 2 & 3 & 4 & 5 & 6 \\ 
        \hline
        1 & -0.012 (0.030) & -0.050 (0.032) & -0.036 (0.030) & -0.025 (0.029) & -0.030 (0.027) & \textbf{-0.078 (0.030)} \\ 
        2 & - & \textbf{-0.086 (0.032)} & \textbf{-0.080 (0.024)} & \textbf{-0.087 (0.028)} & \textbf{-0.088 (0.026)} & \textbf{-0.067 (0.025)} \\ 
        3 & - & - & -0.030 (0.024) & \textbf{-0.054 (0.027)} & -0.048 (0.027) & -0.001 (0.027) \\ 
        4 & - & - & - & 0.006 (0.013) & \textbf{-0.034 (0.015)} & \textbf{-0.060 (0.015)} \\ 
        \hline
    \end{tabular}
    \caption{VoD viewing status}
            \label{tab:outcome3}
    \end{subtable}

    \caption{Point estimates and standard errors of $ATT(g,t)$ for outcome variables. Bold values denote statistical significance at the $\alpha = 0.05$ level. }
    \label{tab:outcomes}
\end{table}

\subsection{Hypothesis tests for the homogeneity of effects}

We next examine whether the causal effects exhibit meaningful heterogeneity along the three dimensions used to summarize $ATT(g,t)$ in Section \ref{sec2.3}. To do so, we apply the proposed bootstrap-based hypothesis testing procedure. Figure \ref{fig:structure_hypo} provides an intuitive overview of the testing structure.

First, for each fixed cohort $g$, we test whether its treatment-effect trajectory is constant over event time. For each outcome of interest, we fix a cohort $g = g^*$ and formulate the null hypothesis as $H_{0, g = g^*}: \ ATT(g^*, g^*) = \cdots = ATT(g^*, T)$.
In our dataset, the specific null hypotheses of interest are as follows:
\begin{align*}
    H_{0, g =1}: & \;\; \tau(1, 1) = \tau(1,2) = \tau(1,3) = \tau(1,4) = \tau(1,5) = \tau(1,6), \\
    H_{0, g =2}: &  \;\;  \tau(2,2) = \tau(2,3) = \tau(2,4) = \tau(2,5) = \tau(2,6), \\
    H_{0, g =3}: & \;\;  \tau(3,3) = \tau(3,4) = \tau(3,5) = \tau(3,6), \\
    H_{0, g =4}: & \;\;  \tau(4,4) = \tau(4,5) = \tau(4,6).
\end{align*}
Second, we assess the cross-cohort homogeneity at a fixed calendar time $t$. Analogously, we test the following hypotheses:
\begin{align*}
    H_{0, t =4}:  \;\; & \tau(1,4) = \tau(2, 4) = \tau(3,4) = \tau(4,4), \\
    H_{0, t =5}:  \;\; & \tau(1,5) = \tau(2, 5) = \tau(3,5) = \tau(4,5), \\
    H_{0, t =6}:  \;\; & \tau(1,6) = \tau(2, 6) = \tau(3,6) = \tau(4,6).
\end{align*}
Moreover, we test whether the $e$-lag effects are constant across different cohorts:
\begin{align*}
    H_{0, e =0}:  \;\; & \tau(1,1) = \tau(2, 2) = \tau(3,3) = \tau(4,4), \\
    H_{0, e =1}:  \;\; & \tau(1,2) = \tau(2, 3) = \tau(3,4) = \tau(4,5), \\
    H_{0, e =2}:  \;\; & \tau(1,3) = \tau(2, 4) = \tau(3,5) = \tau(4,6).
\end{align*}

The p-values from the hypothesis tests are reported in Table \ref{tab:pvals} for all three outcome variables. For total real-time viewing hours, the p-values are generally large, except for the test of $H_{0, g=4}$. Recall from Table \ref{tab:outcome1} that none of the $ATT(g,t)$ estimates for real-time viewing hours were statistically significant at the $\alpha = 0.05$ level. These results suggest that Netflix subscription not only fails to generate a meaningful causal impact on total viewing hours, but also does not produce substantial heterogeneity in the group-time treatment effects.

On the other hand, the outcomes related to VoD content consumption exhibit several sufficiently small p-values indicating statistical significance. For instance, the hypothesis $H_{0, t = 5}$ is rejected at the $\alpha = 0.01$ level for total VoD viewing hours, suggesting that early and late adopters experienced different treatment-effect dynamics in October 2021. Additional heterogeneity is detected for VoD viewing status, including temporal variation within a cohort ($H_{0, g = 4}$), cross-cohort variation at a fixed calendar time ($H_{0, t = 4}$), and differences across $e$-lag effects ($H_{0, e = 0}$). As an illustration, the rejection of $H_{0, e = 0}$ at the $\alpha = 0.05$ level implies that contemporaneous treatment effects exhibit different dynamics depending on the timing of treatment adoption.

Notably, the cohort $g = 4$ and the time period $t = 4$, which correspond to the release of \textit{Squid Game}, do not exhibit a clearly identifiable effect. Although the temporal stability tests for total real-time viewing hours and VoD viewing status within cohort $g = 4$ yield relatively small p-values, the evidence is not strong enough to attribute the patterns directly to the program’s impact. It is plausible that these results reflect a combination of factors—such as seasonal trends or other concurrent influences—rather than the effect of the content release alone.

\begin{table}[!h]
\centering
\begin{tabular}{c | ccc}
\hline
 & Total real-time hours & Total VoD hours & VoD viewing status \\
\hline
$H_{0, g = 1}$ & 0.3126 & 0.0718 & 0.5094 \\
$H_{0, g = 2}$ & 0.1636 & 0.0520 & 0.9296 \\
$H_{0, g = 3}$ & 0.6122 & 0.2960 & 0.1836 \\
$H_{0, g = 4}$ & 0.0734 & 0.5308 & $0.0006^{***}$ \\
\hline
$H_{0, t = 4}$ & 0.9780 & 0.1952 & $0.0056^{**}$ \\
$H_{0, t = 5}$ & 0.6842 & $0.0096^{**}$ & 0.3134 \\
$H_{0, t = 6}$ & 0.2032 & 0.7120 & 0.1386 \\
\hline
$H_{0, e = 0}$ & 0.5948 & 0.0608 & $0.0450^{*}$ \\
$H_{0, e = 1}$ & 0.8806 & 0.1048 & 0.4640 \\
$H_{0, e = 2}$ & 0.8938 & 0.1372 & 0.6278 \\
\hline
\end{tabular}
\caption{P-values from hypothesis tests assessing the homogeneity of $ATT(g,t)$ for the three outcome variables.  Asterisks denote statistical significance at the 0.05 level 
($^{*}: \; p < 0.05$, $^{**}: \; p < 0.01$, $^{***}: \; p < 0.001$). The bootstrap-based tests are conducted with $B = 5{,}000$ iterations.}
\label{tab:pvals}
\end{table}

\section{Discussion}
\label{sec5}

In this study, we developed a novel design-based framework for inference on group-time average treatment effects in time-series settings, equipped with a matching algorithm that constructs the proposed design from observed data. The application of the algorithm to the Netflix–IPTV dataset revealed distinct group-time effects for total IPTV viewing time and VoD consumption behavior, while also allowing a formal assessment of heterogeneity in causal effects across treatment cohorts and timings. Although motivated by this empirical application, the proposed framework is broadly applicable to causal inference problems in longitudinal observational datasets, particularly in settings where treatment timing is potentially confounded by time-varying covariates.

Nevertheless, several venues for future research emerge from this work. First, extending design-based approaches to accommodate more complex treatment dynamics, such as scenarios where units may enter and exit treatment multiple times rather than remaining persistently treated, would significantly broaden the framework’s applicability. Second, when treatment timing is measured on a continuous scale or observed at a fine temporal resolution, the resulting design may become too complex or infeasible due to the limited number of comparable units at each treatment time. An important related direction is to develop approaches regarding temporal aggregation and construct designs that accommodate continuous treatment timings. Finally, although the proposed Reverse-Time Nested Matching algorithm is straightforward to implement by leveraging optimal full matching to reconstruct the nested design, its optimality properties have not yet been explored. Further investigation of the theoretical and computational aspects of the matching algorithm would clarify the performance and guide the development of more precise and efficient estimation procedures. Addressing these challenges would enhance the flexibility and practicality of design-based inference in longitudinal settings.

\bibliographystyle{apalike}

\bibliography{references}
\end{document}